\documentclass[conference]{IEEEtran}

\usepackage{cite}

\ifCLASSINFOpdf
   \usepackage[pdftex]{graphicx}
\else
\fi
\usepackage{graphicx}
\usepackage[cmex10]{amsmath}
\usepackage{amssymb}

\usepackage[tight,footnotesize]{subfigure}

\usepackage{fixltx2e}

\usepackage{stfloats}

\usepackage{url}

\begin{document}
\title{Combined Channel Aggregation and Fragmentation Strategy in Cognitive Radio Networks}
\author{\IEEEauthorblockN{Lei Li\IEEEauthorrefmark{1},
Sihai Zhang\IEEEauthorrefmark{2},
Kaiwei Wang\IEEEauthorrefmark{1} and
Wuyang Zhou\IEEEauthorrefmark{2}}
\IEEEauthorblockA{Wireless Information Network Laboratory\\
University of Science and Technology of China, Hefei, Anhui, P. R. China, 230027\\
Email: \IEEEauthorrefmark{1}\{lyley, wangkw\}@mail.ustc.edu.cn, \IEEEauthorrefmark{2}\{shzhang, wyzhou\}@ustc.edu.cn}
}


\maketitle
\thispagestyle{empty} 
\pagestyle{empty} 

\begin{abstract}
In cognitive radio networks, channel aggregation (CA) and channel fragmentation (CF) techniques have been proposed to enhance the spectrum utilization. While most of the literature studies CA and CF independently, in this paper we combine CA and CF innovatively and present a new spectrum sharing strategy named CAF (Channel Aggregation and Fragmentation). We elaborate on the proposed CAF strategy and derive the balance equation by a continuous time Markov chain (CTMC) model. Then various system performance metrics including blocking probability, dropping probability, spectrum utilization and throughput of the secondary network are evaluated. Both analytical and simulation results show that our strategy lowers the blocking and dropping probabilities and enhances the spectrum utilization and throughput effectively. Moreover, by tuning the bandwidth requirement of each secondary user, different system performance can be achieved.
\end{abstract}


\IEEEpeerreviewmaketitle
\section{Introduction}
In recent years, cognitive radio (CR) \cite{Mitola} has been developed to enable unlicensed secondary users (SUs) to sense and opportunistically access the spectrum temporarily unused by licensed primary users (PUs). By this kind of dynamic spectrum access (DSA), the spectrum utilization can be highly enhanced and spectrum allocation becomes much more flexible than that in traditional constant allocation policies \cite{Akyildiz}. Then channel aggregation (CA) and channel fragmentation (CF) are proposed and widely studied among all of the DSA techniques. CA allows SUs to aggregate multiple contiguous or non-contiguous channels being unused by PUs to support services with higher data rate requirements \cite{JiaoPla,Lee,JiaoLi}. CF, on the other hand, divides a channel to a plurality of sub-channels, thus allowing allocation of a portion of a single channel bandwidth to SUs, which tessellates the granularity of spectrum allocation to fragments or sub-channels \cite{Coffman,Anand}. Both contiguous and non-contiguous CA have been supported by the cognitive radio-based IEEE 802.22 wireless regional area network (WRAN) standard \cite{TV}. For example, up to 3 contiguous TV channels can be bonded to meet the 802.22 data rate requirements. In an enhancement to the standard \cite{Sengupta}, channel carriers are fragmented so that IEEE 802.22 devices can operate over channel bandwidth of 1 to 6 MHz which would allow IEEE 802.22 devices to share the spectrum with incumbent devices such as wireless microphones that only use 1 or 2 MHz of the entire channel assigned.

There has been lots of research work on CA and CF \cite{JiaoPla,Lee,JiaoLi,Coffman,Anand}. In \cite{JiaoPla,Lee}, several CA strategies were proposed, where SUs aggregate constant or variable number of channels. However, as the number of aggregated channels of each SU is restrained from varying during transmission, once the bandwidth requirement can not be satisfied, a new requesting SU will be blocked or an ongoing SU will be dropped. This inevitably leads to the problem of high blocking probability and dropping probability.

To reduce the blocking and dropping probabilities, Jiao \emph{et al.} \cite{JiaoLi} introduced the concept of spectrum adaptation for CA with elastic data traffic, which means an ongoing SU service can adjust the number of aggregated channels according to the availability of channels as well as other SUs' activities. However, the granularity of spectrum allocation is still restricted to channels and thus not sufficiently efficient and flexible. For example, a DTV channel has a bandwidth of as much as 6 MHz \cite{DTV}, which is potentially wide enough for multiple SUs to share some portions of it.

Generally, CF can improve the spectrum utilization by allocating spectrum in a fine-grained manner. However, it's not actually efficient for wideband SUs that require more than one channel to transmit. In \cite{Coffman}, CF was studied theoretically assuming that the fragment size can be arbitrarily small and the main result shows that the average total number of fragments remains bounded. Nevertheless, potentially excessive fragmentation will require highly complex algorithmic solutions for actual spectrum maintenance. Moreover, the loss of orthogonality between spectrum bands due to CF can be exploited by malicious attackers to cause a cognitive service disruption \cite{Anand}. This implies that it is better to restrict the extent of fragmentation appropriately.

The above discussions have motivated us to investigate CA and CF jointly rather than independently. We present a combined channel aggregation and fragmentation (CAF) strategy in which CA and CF are performed adaptively. The benefit of this strategy is two-fold. On one hand, it is inherited from CA which provides a high data rate and improves the spectrum utilization efficiently. On the other hand, CF is allowed in the strategy such that multiple SUs can share the same channel when needed, thus decreasing both the blocking and dropping probabilities. To implement the strategy and for ease of theoretical analysis, we also develop an adaptive spectrum allocation algorithm. Then the system performance in terms of blocking probability, dropping probability, spectrum utilization and throughput of the secondary network are analyzed using a CTMC model and verified by numerical simulations.

The remainder of this paper is organized as follows. In Section II, the system model and assumptions are described as well as our proposed strategy. Section III presents the performance analysis under this strategy. Numerical results and discussions are given in Section IV, while conclusions are drawn in Section V.

\section{System Model}
\subsection{Basic Model and Assumptions}
Consider an overlay CR network in which a secondary base station (BS) manages its own cell and all associated SUs. When a new SU initiates a service request, the BS decides whether to allocate radio resources (typically bandwidth in this paper) to the SU. If enough resources are unavailable, the SU will be blocked. The allocation result can be delivered on the channel monitored by the specific SU which has been obtained through the association procedures \cite{Cordeiro}, or on a predefined common control channel \cite{Song}. The BS shall instruct its associated SUs to perform periodic spectrum sensing to update the spectrum occupancy map that identifies for each channel whether PUs have been detected or not \cite{Cordeiro}. As the sensing duration is much more shorter than the transmission duration, we ignore its overhead in this paper.

In the CR network, a total number of $N$ channels are shared by PUs and SUs and each channel is assumed to have a normalized unit bandwidth. Every PU occupies only one channel while an SU may aggregate multiple channels by CA or even utilize some portions of channels by CF for a service transmission, i.e., the elastic data traffic is considered. Suppose the SUs are homogenous and every SU has the same bandwidth requirement of $[B_{m},B_{M}]$, where $B_{m}$ and $B_{M}$ represent the minimum and maximum bandwidth requirements respectively. Without loss of generality, we assume that $B_{m}\geq 1$. However, our results can be generalized to $B_{m}<1$.

The arrivals of PUs and SUs are assumed to be independent Poisson processes with arrival rates $\lambda_{p}$ and $\lambda_{s}$ respectively. The service time on one channel or one unit bandwidth follows exponential distributions with mean $1/\mu_{p}$ and $1/\mu_{s}$ respectively. The SU service rate is composed of two parts, i.e., $\mu_{s}=h_{s}+r_{s}$, where $1/h_{s}$ represents the mean channel holding time while $1/r_{s}$ is the mean cell residence time, considering that the service time is the minimum of the channel holding time and cell residence time who are also assumed to be exponentially distributed \cite{Tang}.

\subsection{CAF Strategy}
We propose a combined channel aggregation and fragmentation (CAF) strategy. In CAF, SUs have the capability to dynamically adjust their currently used spectrum according to the system load and channel availability. Specifically, when there is only a low system load and enough idle channels are available, the SUs occupy the spectrum by CA to maximize the service rate; otherwise, the channels are fragmented such that multiple SUs can share them. In CAF strategy, CA and CF are performed adaptively. While CA provides a high data rate and utilizes the spectrum efficiently, CF guarantees the blocking and dropping probabilities being as low as possible. By CAF, the granularity of spectrum allocation is properly extended from channels to sub-channels or fragments when needed, which ensures the fragmentation will not be excessive.

\begin{figure}[!t]
    \centering
    \includegraphics[width=0.48\textwidth]{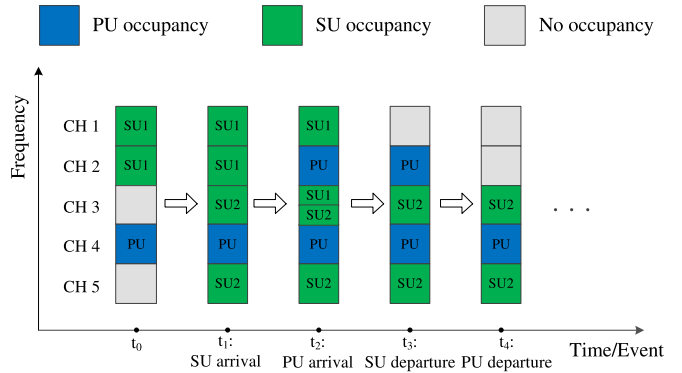}
    \caption{An example of the procedures of CAF, where $N=5$, $B_{m}=1$ and $B_{M}=2$.}
    \label{procedure}
\end{figure}

To implement the strategy from a more comprehensive perspective, we also present an adaptive spectrum allocation algorithm named \emph{Equal Sharing Algorithm}. The algorithm distributes available spectrum among ongoing or new coming SUs equally\footnote{This equal sharing process has been widely used in various wireless communication networks for resource allocation \cite{Jeng,Perez,Ureta}. However, to the best of our knowledge, it has not been introduced to CR networks before.}, given that the bandwidth requirement of each SU is not violated. As we consider homogenous SUs with elastic data traffic, the equal sharing process guarantees the fairness among different SUs and provides much flexibility for spectrum allocation. From another point of view, the algorithm greatly facilitates the theoretical analysis of CAF strategy besides simulations.

Basically, there are three scenarios where Equal Sharing Algorithm works. When a new SU arrives with enough idle channels unavailable, the equal sharing process between the new SU and other ongoing SUs helps to reduce the blocking probability. Similarly, when an ongoing SU is obliged to handoff but no idle channels exist, the equal sharing process among ongoing SUs decreases the dropping probability. And when some spectrum is newly vacated, re-occupancy of it by ongoing SUs can promote the spectrum utilization and data rate of each SU.

In summary, Equal Sharing Algorithm is executed whenever the maximum bandwidth requirement $B_{M}$ cannot be fulfilled for all SUs; otherwise, every SU will utilize a maximum bandwidth and transmit with a maximum service rate. The detailed procedures of CAF strategy are listed below and shown in Fig. \ref{procedure} as an example.
\begin{enumerate}
  \item When a new SU arrives, it attempts to occupy the maximum bandwidth $B_{M}$ by CA if there exist enough idle channels in the system; otherwise, the available channels are fragmented and equally shared by ongoing SUs plus this new SU by Equal Sharing Algorithm, ensuring that every SU (both the ongoing and new coming ones) keeps a bandwidth not less than $B_{m}$, otherwise the new SU will be blocked.
  \item When a new PU arrives, it randomly chooses a channel not used by other PUs to access. If this particular channel is occupied by SU(s), the latter will either handoff to another idle channel if there is still one, or share the available spectrum by Equal Sharing Algorithm, ensuring that every SU keeps a bandwidth not less than $B_{m}$, otherwise the specific SU(s) will be dropped.
  \item Once a PU or SU completes transmission and leaves the system, the residual SUs will equally share the vacant spectrum by Equal Sharing Algorithm promptly, given that the maximum bandwidth requirement is not violated; that is, no SU could ever occupy a bandwidth larger than $B_{M}$.
\end{enumerate}

\begin{figure}[!t]
    \centering
    \includegraphics[height=0.35\textwidth]{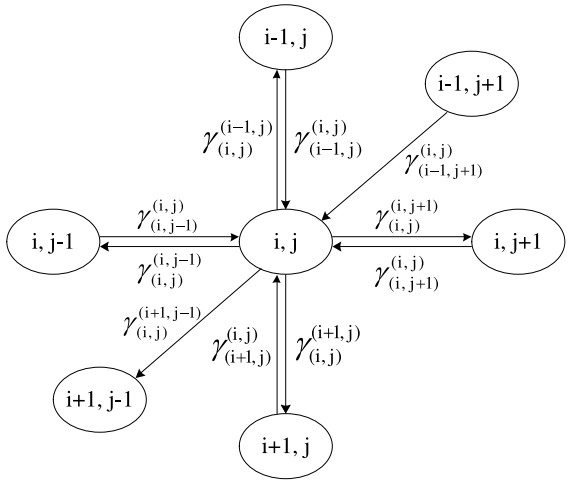}
    \caption{The state transition diagram in CAF strategy.}
    \label{st}
\end{figure}

\textbf{\emph{Remark 1:}} In CAF, when a spectrum rearrangement triggered by Equal Sharing Algorithm is needed, the results of the rearrangement can be conveyed on the common control channel. Alternatively, the results can be inserted into the common frame header shared by the aggregated channels \cite{Cordeiro,Sengupta} and broadcast to all the SUs involved to inform them about the spectrum to be used in the next frame. Herein, one particular advantage of CA is highlighted, i.e. even when one of the aggregated channels is preempted by a PU, the availability of the residual aggregated channels can still guarantee the successful communication between the SU and BS. Therefore, the algorithm can be assumed to be implemented effectively. Moreover, as the cognitive capability of an SU builds upon the software-defined radio (SDR) \cite{Mitola}, its operating parameters such as frequency band, code and modulation, and transmit power, etc. can be reconfigured flexibly such that the efficiency of the rearrangement is provisioned.

\textbf{\emph{Remark 2:}} The maximum number of fragments in CAF is bounded due to the limited number of SUs. Specifically, since a maximum number $\lfloor\frac{N}{B_{m}}\rfloor$ of SUs are allowed in the network and at most one sub-channel (fragment) besides the aggregated channels can be occupied by each SU, the maximum number of fragments in CAF is restricted to $\lfloor\frac{N}{B_{m}}\rfloor$.

\section{Performance Analysis}
In this section, we analyze the system performance of the secondary network in terms of various metrics under our proposed CAF strategy by using a continuous time Markov chain (CTMC) model.

\subsection{CTMC Model}
In the CTMC model, each state can be represented as an integer pair $(i,j)$, where $i$ is the number of PUs and $j$ is the number of SUs in the system. The feasible state space is $S=\{(i,j)|0\leq i\leq N,0\leq j\leq\lfloor\frac{N}{B_{m}}\rfloor,i+jB_{m}\leq N\}$, where $\lfloor x\rfloor$ is the maximum integer less than or equal to $x$. Note that there will be no more than $\lfloor\frac{N}{B_{m}}\rfloor$ SUs allowed in the system because each SU has a minimum bandwidth requirement of $B_{m}$. Then the bandwidth and service rate of each SU in state $(i,j)$ are
\begin{equation}\label{1}
    B(i,j)=\left\{\begin{array}{ll}
    \min\{B_{M},\max\{B_{m},\frac{N-i}{j}\}\},\\
    \qquad\textrm{if $0\leq i\leq\lfloor N-B_{m}\rfloor,1\leq j\leq\lfloor\frac{N}{B_{m}}\rfloor$;}\\
    0, \quad \textrm{otherwise.}
    \end{array}\right.
\end{equation}
and
\begin{equation}\label{2}
    \mu_{s}(i,j)=\left\{\begin{array}{ll}
    B(i,j)\cdot h_{s}+r_{s},\\
    \qquad\textrm{if $0\leq i\leq\lfloor N-B_{m}\rfloor,1\leq j\leq\lfloor\frac{N}{B_{m}}\rfloor$;}\\
    0, \quad \textrm{otherwise.}
    \end{array}\right.
\end{equation}
where the mean residence time $1/r_{s}$ is independent of the bandwidth an SU occupies.

\begin{figure}[!t]
    \centering
    \includegraphics[height=0.35\textwidth]{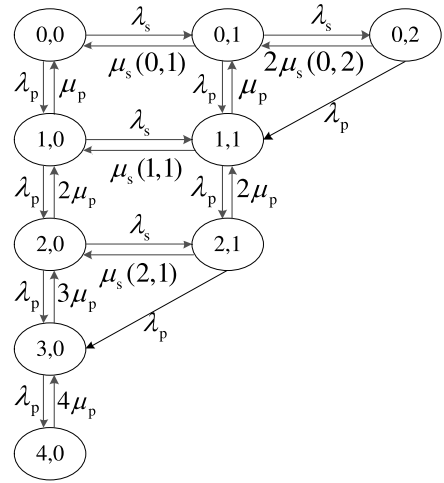}
    \caption{An example of the state transition diagram in CAF strategy when $N=4$, $B_{m}=2$ and $B_{M}=3$.}
    \label{egst}
 \end{figure}

The state transition diagram is depicted in Fig. \ref{st}. The notation $\gamma_{(i,j)}^{(i',j')}$ represents the transition rate from one feasible state $(i,j)$ to another feasible state $(i',j')$. The transitions from $(i,j)$ to all other feasible states are summarized as follows:
\begin{equation}\label{3}
    \textrm{Normal cases: }\left\{\begin{aligned}
    \gamma_{(i,j)}^{(i,j-1)}&=j\mu_{s}(i,j);\\
    \gamma_{(i,j)}^{(i,j+1)}&=\lambda_{s};\\
    \gamma_{(i,j)}^{(i-1,j)}&=i\mu_{p};\\
    \gamma_{(i,j)}^{(i+1,j)}&=\lambda_{p},~\textrm{if $i+1+jB_{m}\leq N$.}
    \end{aligned}\right.
\end{equation}
\begin{equation}\label{4}
    \textrm{Dropping case: }\gamma_{(i,j)}^{(i+1,j-1)}=\lambda_{p},~\textrm{if $i+1+jB_{m}>N$;}
\end{equation}
\begin{equation}\label{5}
    \textrm{Blocking case: }\gamma_{(i,j)}^{(i,j)}=\lambda_{s},~\textrm{if $i+(j+1)B_{m}>N$.}
\end{equation}

Normal cases in (\ref{3}) characterize the normal arrivals or departures of PUs and SUs respectively. For example, $\gamma_{(i,j)}^{(i,j-1)}$ is the transition rate from a feasible state $(i,j)$ to $(i,j-1)$, which is caused by the departure of an ongoing SU from the system. A dropping event happens when a new PU arrives and at least one SU can no longer be served even after the equal sharing process, which corresponds to (\ref{4}). An SU blocking event is caused by insufficient spectrum resources upon arrival of a new SU even after the equal sharing process corresponding to (\ref{5}). Note that all states involved above must be feasible; otherwise, the corresponding transition rate is set to zero. An example of the state transition diagram when $N=4$, $B_{m}=2$ and $B_{M}=3$ is depicted in Fig. \ref{egst}.

Let $\pi(i,j)$ be the steady state probability when in state $(i,j)$. A feasible state indication function is defined as
$I(i,j)=\left\{\begin{array}{ll}
1, & \textrm{if $(i,j)\in S$}\\
0, & \textrm{otherwise}
\end{array}\right.$.
Then, we can describe the balance equation as
\begin{equation}\label{6}
\begin{aligned}
    &\underbrace{\sum\limits_{i'=0}^{N}\sum\limits_{j'=0}^{\lfloor\frac{N}{B_{m}}\rfloor}}\limits_{(i',j')\neq (i,j)}\pi(i,j)\gamma_{(i,j)}^{(i',j')}I(i,j)I(i',j')=\\
    &\underbrace{\sum\limits_{i'=0}^{N}\sum\limits_{j'=0}^{\lfloor\frac{N}{B_{m}}\rfloor}}\limits_{(i',j')\neq (i,j)}\pi(i',j')\gamma_{(i',j')}^{(i,j)}I(i,j)I(i',j'),
\end{aligned}
\end{equation}
where $0\leq i\leq N,0\leq j\leq\lfloor\frac{N}{B_{m}}\rfloor$. Additionally, the normalization condition is
\begin{equation}\label{7}
    \sum\limits_{i=0}^{N}\sum\limits_{j=0}^{\lfloor\frac{N}{B_{m}}\rfloor}\pi(i,j)I(i,j)=1.
\end{equation}

Combining (\ref{6}) and (\ref{7}), the balance equation can be solved by using the same method in \cite{Xing}. Briefly speaking, the method constructs an infinitesimal generator matrix whose elements are composed of the transition rates between different states. Then by putting the conditions (\ref{6}) and (\ref{7}) into a compact matrix equation and using the minimum mean-square error (MMSE) criterion, all the steady state probabilities can be uniquely solved. This method is very simple but efficient. Once the steady state probabilities are obtained, various metrics that characterize the system performance can be calculated in the following.

\subsection{Blocking Probability}
An SU blocking event happens when there are insufficient spectrum resources upon arrival of a new SU even after the equal sharing process. Let $P_{b}$ denote the SU blocking probability. It's characterized by the blocking rate $\gamma_{(i,j)}^{(i,j)}$ and SU arrival rate $\lambda_{s}$, i.e.,
\begin{equation}\label{9}
\begin{aligned}
    P_{b}&=\frac{1}{\lambda_{s}}\sum\limits_{i=0}^{N}\sum\limits_{j=0}^{\lfloor\frac{N}{B_{m}}\rfloor} \gamma_{(i,j)}^{(i,j)}\pi(i,j)I(i,j)\\
         &=\sum\limits_{i=0}^{N}\sum\limits_{j=\lceil\frac{N-i}{B_{m}}\rceil}^{\lfloor\frac{N}{B_{m}}\rfloor} \pi(i,j)I(i,j).
\end{aligned}
\end{equation}

\subsection{Dropping Probability}
A dropping event happens when a new PU arrives and at least one SU can no longer be served even after the equal sharing process. The SU dropping probability $P_{d}$ can be expressed as the mean dropping rate divided by the mean admitted SUs rate, that is
\begin{equation}\label{10}
\begin{aligned}
    P_{d}=\frac{1}{(1-P_{b})\lambda_{s}}\sum\limits_{i=0}^{N}\sum\limits_{j=0}^{\lfloor\frac{N}{B_{m}}\rfloor}
    \gamma_{(i,j)}^{(i+1,j-1)} \\
    \cdot\pi(i,j)I(i,j)I(i+1,j-1).
\end{aligned}
\end{equation}

\subsection{Spectrum Utilization}
The spectrum utilization is defined as the ratio of the mean channel occupancy by SUs to the total channel bandwidth. Considering that each channel is assumed to have a unit bandwidth, the spectrum utilization can then be calculated by
\begin{equation}\label{11}
    \eta=\frac{1}{N}\sum\limits_{i=0}^{N}\sum\limits_{j=0}^{\lfloor\frac{N}{B_{m}}\rfloor} jB(i,j)\pi(i,j)I(i,j).
\end{equation}

\subsection{Throughput}
The throughput of the secondary network, denoted by $Th_{SU}$, is defined as the mean number of service completions per unit time \cite{Ngatched}. Thus,
\begin{equation}\label{12}
    Th_{SU}=\sum\limits_{i=0}^{N}\sum\limits_{j=0}^{\lfloor\frac{N}{B_{m}}\rfloor}
    \pi(i,j)I(i,j)j\mu_{s}(i,j).
\end{equation}

\section{Numerical Result}
We present both analytical and simulation results to evaluate the system performance. The CCA strategy \cite{Lee} is adopted for comparison, which is actually a special case of CAF when $B_{M}=B_{m}$. In CCA, each SU aggregates a constant number of channels, i.e. the fragmentation is not allowed. The basic simulation parameters are set as $N=12$, $\mu_{p}=0.45$, $\lambda_{s}=7.2$, $h_{s}=1$ and $r_{s}=1$. We use MATLAB to directly solve the balance equations to obtain the analytical results of various performance metrics. Simulations are executed by generating both PU and SU services according to the assumed distributions using C program. A total number of $10^6$ is set as the upper bound of iteration and 100 realizations are averaged to obtain the final results. In all the figures, the solid lines are the analytical results while the marks are the simulation results.

\begin{figure}[!t]
    \centering
    \includegraphics[width=0.48\textwidth]{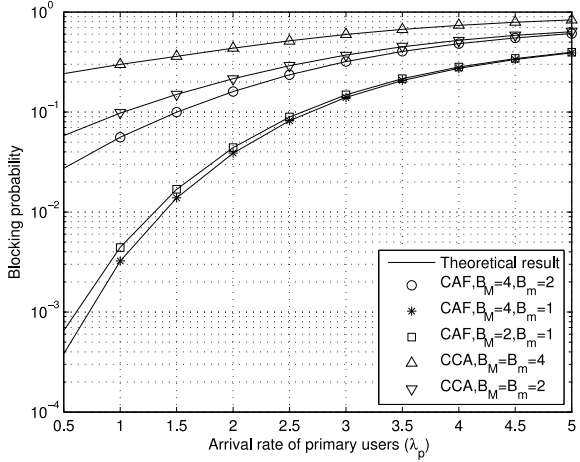}
    \caption{Blocking probability of SUs as a function of $\lambda_{p}$.}
    \label{b1}
\end{figure}
\begin{figure}[!t]
    \centering
    \includegraphics[width=0.48\textwidth]{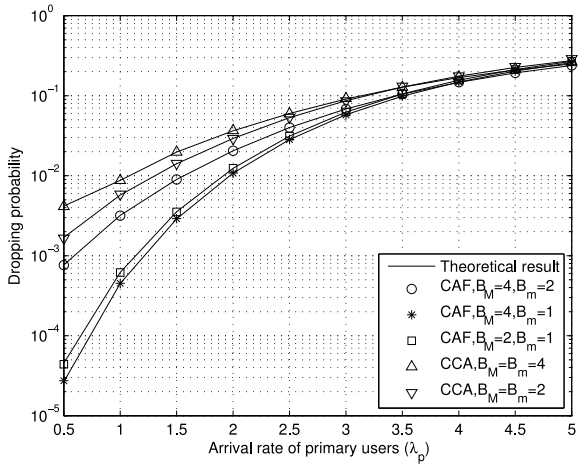}
    \caption{Dropping probability of SUs as a function of $\lambda_{p}$.}
    \label{b2}
\end{figure}

\begin{figure}[!t]
    \centering
    \includegraphics[width=0.48\textwidth]{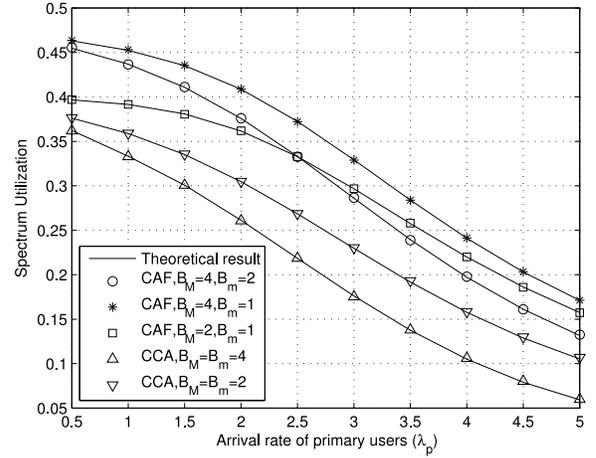}
    \caption{Spectrum utilization as a function of $\lambda_{p}$.}
    \label{b3}
\end{figure}

\begin{figure}[!t]
    \centering
    \includegraphics[width=0.48\textwidth]{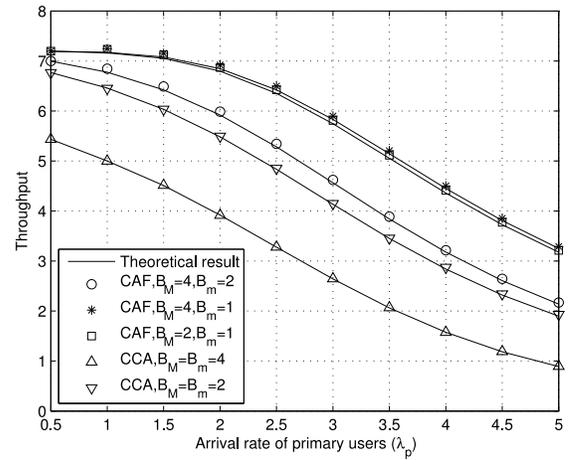}
    \caption{Throughput of the secondary network as a function of $\lambda_{p}$.}
    \label{b4}
\end{figure}

We compare both the maximum aggregation rule in which each SU always aggregates the maximum number $B_{M}$ of channels, and the minimum aggregation rule in which each SU always aggregates the minimum number $B_{m}$ of channels. These two rules represent the lower and upper capacity\footnote{Herein, the capacity refers to the maximum number of SUs that can be accommodated in the secondary network.} limits of the CCA strategy respectively. Intuitively, our proposed CAF strategy seems to be a compromise of these two rules such that the system performance shall fall in between them. Interestingly, however, the simulations disprove this intuition. From Figs. \ref{b1}-\ref{b4}, we can observe that the CAF strategy (corresponding to CAF with $B_{M}=4$, $B_{m}=2$) outperforms both the maximum aggregation rule (corresponding to CCA with $B_{M}=B_{m}=4$) and the minimum aggregation rule (corresponding to CCA with $B_{M}=B_{m}=2$) in terms of blocking probability, dropping probability, spectrum utilization and throughput of the secondary network. It can be explained like this. On one hand, in the CAF strategy each SU always attempts to aggregate the maximum number of channels greedily, just like the maximum aggregation rule. On the other hand, variable rather than constant minimum bandwidth enables the SUs to transmit with larger average data rates such that their services can be completed in a shorter time, which leads to a lower dropping probability for ongoing SUs and provides more service chances for new coming SUs.


Then, the impacts of different bandwidth requirements of SUs on the performance of the CAF strategy are investigated.  Fig. \ref{b1} shows the SU blocking probability as a function of $\lambda_{p}$ with different $B_{M}$ and $B_{m}$. It's observed that by reducing $B_{m}$, the SU blocking probability can be greatly lowered. This is reasonable considering the fact that with a smaller $B_{m}$, more SUs can be accommodated in the system, thus less blocking events will happen. Besides, the blocking probability decreases slightly when $B_{M}$ increases. It is because the blocking probability is mainly affected by $B_{m}$, which determines the maximum number of SUs admitted in the system. However, with a bigger $B_{M}$, ongoing SUs can obtain higher average service rates and experience less average service time, which creates more chances for new coming SUs to be served and thus slightly lowers the blocking probability.

Fig. \ref{b2} illustrates the variation of SU dropping probability as $\lambda_{p}$ increases. A similar phenomenon is observed as that in Fig. \ref{b1}. By reducing $B_{m}$, the SU dropping probability can be lowered to a large extent; while it decreases slightly when $B_{M}$ increases. Actually, with a bigger $B_{M}$, an ongoing SU can obtain a higher average service rate, which increases the probability that the service completes without being disrupted by PUs and thus slightly lowers the dropping probability.

Fig. \ref{b3} shows the spectrum utilization under different $B_{M}$ and $B_{m}$. It is observed that when $\lambda_{p}$ is relatively low, increasing $B_{M}$ improves the spectrum utilization more effectively than reducing $B_{m}$; while it's inverse when $\lambda_{p}$ is high. The reason is that the SU's activity plays a main role when $\lambda_{p}$ is low, thus the spectrum can be utilized more efficiently with a larger $B_{M}$. However, when $\lambda_{p}$ is high, the PU's activity plays a main role. Thus by reducing $B_{m}$, more SUs can be accommodated in the system.

Fig. \ref{b4} shows how the throughput of the secondary network varies with different $B_{M}$ and $B_{m}$ as $\lambda_{p}$ increases. We can observe that different throughput can be achieved by tuning $B_{M}$ and $B_{m}$.

\section{Conclusion}
In this paper, an innovative spectrum sharing strategy which combines channel aggregation and channel fragmentation in CR networks is proposed and investigated by using a CTMC model. The system performance in terms of various metrics is evaluated and compared through both mathematical analysis and simulations. Numerical results show that our proposed strategy lowers the SU blocking and dropping probabilities and enhances the spectrum utilization and throughput of the secondary network effectively. Moreover, by tuning the lower and upper bounds of the bandwidth requirement of each SU, different system performance can be achieved. Our future work is to introduce a queuing scheme to our strategy and reassess the system performance.

\section*{Acknowledgment}
This work is partially supported by the National Major Special Projects in Science and Technology of China under grant  2010ZX03003-001, 2010ZX03005-003, 2011ZX03003-003-04.


\end{document}